\documentclass[10pt,journal,compsoc]{IEEEtran}

\ifCLASSOPTIONcompsoc
  \usepackage[nocompress]{cite}
\else
  \usepackage{cite}
\fi

\ifCLASSINFOpdf

\else

\fi

\usepackage{graphicx}
\usepackage{subcaption}
\usepackage{color}
\usepackage{xcolor}
\usepackage{diagbox}
\usepackage{amssymb}
\usepackage{bbding}
\usepackage{amsmath}
\usepackage{bm}
\usepackage{hyperref}
\usepackage{booktabs}
\usepackage{multirow}
\usepackage{tabularx}
\usepackage{tikz}
\usepackage{colortbl}

\usepackage{xcolor}
\renewcommand{\color}[1]{}

\begin{document}

\title{DASKT: A Dynamic Affect  Simulation Method for Knowledge Tracing }

\author{Xinjie~Sun,
        Kai~Zhang,~\IEEEmembership{Member,~IEEE,}
        Qi~Liu,~\IEEEmembership{Member,~IEEE,}
        Shuanghong~Shen,
        Fei~Wang,
        Yuxiang~Guo,
        Enhong~Chen,~\IEEEmembership{Fellow,~IEEE,}
       \IEEEcompsocitemizethanks{
       \IEEEcompsocthanksitem Xinjie Sun, Shuanghong Shen,  Fei Wang and Yuxiang Guo, are with the School of Computer Science and Technology, University of Science and Technology of China, State Key Laboratory of Cognitive Intelligence, Hefei, Anhui 230027, China.\protect \\
E-mail:  \{xinjiesun, closer, wf314159, guoyx18\}@mail.ustc.edu.cn

       \IEEEcompsocthanksitem    Kai Zhang, Qi Liu and Enhong Chen are with the State Key Laboratory of Cognitive Intelligence ,  School of Computer Science and Technology, University of Science and Technology of China, Hefei, Anhui 230027, China.\protect \
E-mail:  \{kkzhang08, qiliuql, cheneh\}@ustc.edu.cn
\IEEEcompsocthanksitem    Xinjie Sun  is with  Liupanshui normal university, Liupanshui, Guizhou 553004, China.\protect \

}

\thanks{
Corresponding author:  Kai~Zhang.}
}

\markboth{Journal of \LaTeX\ Class Files}%
{Shell \MakeLowercase{\textit{et al.}}: Bare Demo of IEEEtran.cls for Computer Society Journals}

\IEEEtitleabstractindextext{%
\begin{abstract}
Knowledge Tracing (KT) predicts future performance by modeling students' historical interactions, and understanding students' affective  states can enhance the effectiveness of KT, thereby improving the quality of education. Although  traditional  KT values students' cognition and learning behaviors, efficient evaluation of students' affective  states and their application in KT still require further exploration due to the non-affect-oriented   nature of the data and budget constraints. To address this issue, we propose a computation-driven approach, \textbf{D}ynamic \textbf{A}ffect \textbf{S}imulation \textbf{K}nowledge \textbf{T}racing (DASKT), to explore the impact of various student affective  states (such as frustration, concentration, boredom, and confusion) on their knowledge states. In this model, we first extract affective factors from students' non-affect-oriented behavioral  data, then use clustering and spatiotemporal sequence modeling to accurately simulate students' dynamic affect changes when dealing with different problems. Subsequently, {\color{blue}we incorporate affect with time-series analysis to improve the model's ability to infer knowledge states over time and space.} Extensive experimental results on two public real-world educational datasets show that DASKT can achieve more reasonable knowledge states under the effect of students' affective  states. Moreover, DASKT outperforms the most advanced KT methods in predicting student performance. Our research highlights a promising avenue for future KT studies, focusing on achieving high interpretability and accuracy.

\end{abstract}

}

\maketitle
\IEEEdisplaynontitleabstractindextext
\IEEEpeerreviewmaketitle

\begin{figure}
    \centering
    \includegraphics[width=0.5\textwidth]{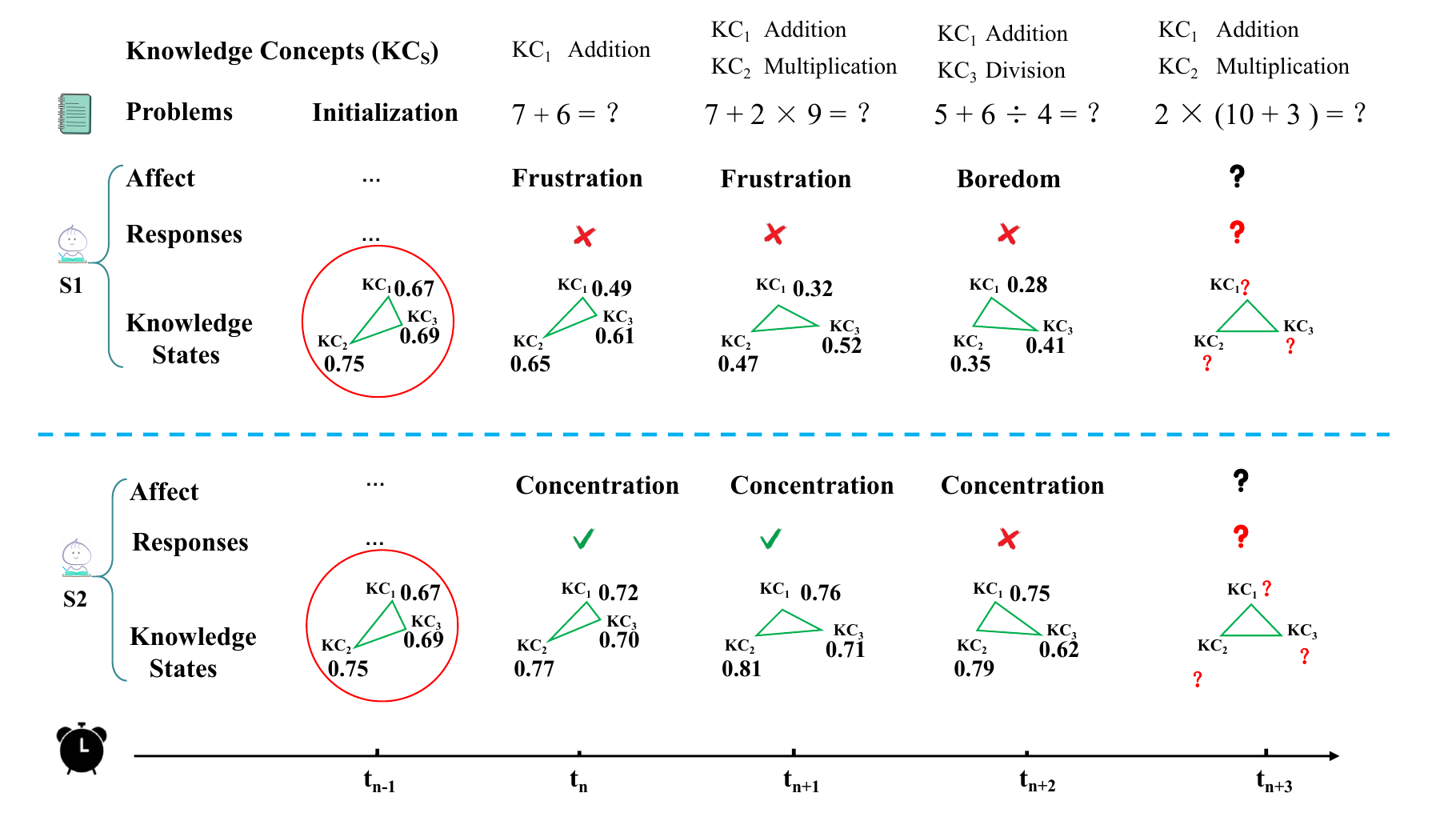}
    \caption{\textbf{A toy example illustrating the mastery level of three knowledge concepts in the same exercise sequence for two students who have identical cognitive abilities at the initial stage. Student $s1$ experiences affect of frustration and boredom throughout the learning process, while Student $s2$ consistently maintains a state of concentration.}\label{figure1}}
\end{figure}

\IEEEraisesectionheading{\section{Introduction}\label{sec:introduction}}
\IEEEPARstart{K}{nowledge} Tracing (KT) is a task that dynamically estimates students' knowledge states (e.g., mastery of Addition, Multiplication) over time, using their previous learning interactions (e.g., responses to a series of exercise questions). This allows for the prediction of student performance on the next question\cite{bib034,bib025}(illustrated in Figure \ref{figure1}). Known as a foundational task in the field of intelligent education, the wide application of KT in online teaching platforms, interactive learning environments, and intelligent tutoring systems \cite{bib036} demonstrates its profound academic and practical impact.

In recent years, deep neural networks have made rapid progress, and a series of deep learning based knowledge tracing (DLKT) models have emerged, such as deep sequence KT models based on autoregression \cite{bib019,bib011,bib025,bib027} to simulate students' complex learning behaviors. To further analyze students' learning trajectories, the field of  has introduced attention mechanisms \cite{bib021,bib034,bib035,bib028}. Graph KT models \cite{bib029,bib019,bib024} use knowledge concepts (KCs) and students' learning paths as nodes and edges of the graph to simulate the learning process. To compensate for the weak interpretability of neural networks, more and more researchers have recently incorporated educational theory into their approaches, as evidenced by studies \cite{bib039,bib040,bib041,bib042}, increasing the interpretability of  KT. {\color{blue} Moreover, using affect detectors for student affect detection and integrating high-quality student affect into KT models can enhance model predictive performance \cite{bib059}. These innovative methods have significantly promoted the refinement and dynamization of the  task.}

Although DLKT methods offer a new perspective for solving KT questions, the degree to which students master knowledge states still largely depends directly on student interactions with the questions \cite{bib046,bib047}. Yet, an over-reliance on interaction data for evaluating student knowledge states, while neglecting individual student differences, often leads to a mechanistic DLKT model. Figure \ref{figure1} illustrates the mastery level of the same knowledge 
 concept (KC) for two students. Specifically, the mastery levels of three knowledge concepts (Addition, Multiplication, Division) by Student $s1$ and Student $s2$ are the same before time step $t_{n}$, suggesting their cognitive abilities are equivalent at this moment. Despite their subsequent learning behavior being the same, their responses to the problems are significantly different, leading to substantial differences in their knowledge states. This situation complicates understanding student performance solely from the mastery of knowledge states, thereby diminishing the interpretability of the model. Upon closer inspection of Figure \ref{figure1}, one might find that this difference could be due to their varying affective  states. Consequently, it is essential to use student exercise records to simulate student affect during problem-solving and to estimate knowledge states. This paper delves into the precise computation of affect coherence simulation in the absence of costly affect labels, while thoroughly exploring the impact of affect on knowledge states.

Students' learning histories and their interactions with intelligent tutoring systems vary from person to person. In order to accurately simulate students' affective  states when facing different problems, and to incorporate these affective  states into  tasks, it is crucial to conduct personalized and precise assessments of students' affect and mastery of knowledge states. However, incorporating affective  states into the  process faces three important challenges. \textbf{Firstly}, we need to identify and validate affective factors. Traditional affect assessments mainly rely on on-site tracking assessments or the use of a large amount of hardware and software equipment to collect students' text, expressions, physiological signals and other affect expression factors, to develop affect detectors. However, this method requires a significant amount of time and financial cost in the implementation process, posing significant limitations for large-scale application and promotion. \textbf{Secondly}, we need to simulate students' affect changes based on these affective factors. Human affect changes are highly continuous, while the affective factors obtained based on learning exercise records are discrete, making it difficult for us to accurately assess students' affect changes. \textbf{Lastly}, we need to explore how to carry out effective knowledge states assessments under affective states. Specifically, how to integrate affective states  with DLKT to enhance the quality of model inference, and how to achieve the unification of affect understanding and complex inference models. Solving these problems requires a delicate operation, that is, to take into consideration both the personalized affect of students and their understanding of knowledge, to ensure a more objective assessment of students' knowledge states.

To address these challenges, we propose a computation-driven dynamic affect simulation knowledge tracing (DASKT) model. This is an innovative  method in which students' affective  states play a significant role in assessing students' knowledge states. The model contains several key innovations aimed at enhancing the effectiveness and interpretability of KT.  \textbf{Firstly},  in order to identify affective factors within behavioral data that is not primarily oriented towards affect, we have categorized these affective factors into three key dimensions: students' confidence, interest, and effort. We use accuracy and response time to simulate confidence, the frequency of participation in exercises to simulate interest, and the number of attempts and response intervals to simulate effort. \textbf{Secondly}, to simulate students' dynamic affect changes, we adopted a time-series-based affect computation model and performed deep spatial state capture on its generated affective  states to obtain the spatiotemporal continuity features of affect. \textbf{Finally}, to integrate affective states with DLKT, we adopted a sequence modeling approach under affective  states. This method enhances the representation of student interactions and can infer students' knowledge states at a high quality.

In summary, this is our solution to the aforementioned challenges, and we hope that through this innovative method, we can more accurately assess and track students' knowledge states. The main contributions of this research can be summarized as follows:

\begin{itemize}	
\item \textbf{Proposed}: A novel affect recognition method extracts affective factors from students' non-affect-oriented behavioral data, reducing reliance on costly affect labels. This innovative approach could inspire new research directions and broaden application scenarios.

\item \textbf{Established}: A dynamic affect simulation framework is introduced, enabling the capture and simulation of students' affect changes\footnote{To encourage reproducible research, we make the data and code available at https://github.com/xinjiesun-ustc/DASKT to ensure accessibility and promote further innovation in this field.}. This method enhances understanding and prediction of students' learning behaviors and knowledge mastery states.

\item  \textbf{Integrated}: Affective states is effectively combined with DLKT for the first time. Experimental results demonstrate the potential for improving inference quality and interpretability of KT models, with implications for educational technology.
\end{itemize}

\section{Related work}
Accurately simulating the various affect  that students exhibit during the exercise process, assessing their affect  when facing different problems, and effectively integrating these affect  with students' knowledge states are undoubtedly tasks of great value. Subsequently, we will introduce related research work from two perspectives: knowledge tracing and affective states recognition methods.

\subsection{Konwledge Tracing}

Recently, deep learning technologies have been widely applied to Knowledge Tracing (KT) tasks for modeling students' historical learning and predicting future performance. Existing DLKT methods can be divided into five categories: C1: Improving the basic structure of the model; C2: Adding problem-solving features; C3: Simulating educational theory; C4: Based on graph structure; C5: Other networks structure.

\vspace{0.1cm}

\noindent \textbf{C1: Improving the basic structure of the model.} In the realm of Knowledge Tracing (KT), research focuses on refining model structures to monitor students' knowledge mastery accurately. Deep Knowledge Tracing (DKT) \cite{bib019} introduced deep learning, using LSTM layers to assess student understanding. Memory networks, like those in \cite{bib031, bib032, bib033}, differentiate mastery levels across knowledge domains by storing and retrieving concepts. Attention mechanisms prioritize relevant data, enhancing information extraction in learning processes \cite{bib034, bib035, bib036}. These advances enhance tracing accuracy and deepen insights into student learning.

\vspace{0.1cm}

\noindent \textbf{C2: Adding problem-solving features.} Researchers are integrating problem-solving features to better understand and predict students' knowledge mastery. Cheung, et al. \cite{bib037} predicted and actual responses are encoded as binary bits and combined with original features for LSTM input. Meanwhile, Sonkar, et al. \cite{bib038} introduce problem IDs to predict learners' success probabilities on individual problems, adjusting predictions to address the assumption that all problems are equivalent observations of abilities.
\vspace{0.1cm}

\noindent \textbf{C3: Simulating educational theory (**).}  {\color{blue}To enhance the interpretability of KT, researchers are integrating educational theories into DKT.  Wang, et al. \cite{bib039}  introduced a more interpretable cognitive diagnosis method into DKT. Moreover, some studies explore enhancing interpretability by simulating learning gains and forgetting behaviors \cite{bib025, bib040}, considering difficulty effects on knowledge \cite{bib041}, and modeling exercise relationships \cite{bib042}. \textbf{Our DASKT model is particularly relevant here, as it effectively uses student affect to influence learning outcomes.}}

\vspace{0.1cm}

\noindent \textbf{C4:  Based on graph structure (*).} DLKT models are incorporating graph learning technologies like Graph Neural Networks (GNNs) for enhanced representation capabilities.  Nakagawa, et al.\cite{bib029} integrated GNNs into KT tasks with significant success. Other studies exploit graph structures to model relationships between problems and skills effectively \cite{bib019,  bib046, bib024}. \textbf{Our DASKT model similarly employs Graph Networks to capture the evolution of affect changes over time, diverging from using them between KCs.}

\vspace{0.1cm}
\noindent \textbf{C5: Other models' structure.}  DLKT  model problem IDs, KCs, and their interaction relationships using  deep learning technologies such as graph attention networks, text features, and adversarial techniques. For example, Pandey, et al.\cite{bib047} uses attention mechanisms to model the relationship between KC; Liu, et al.\cite{bib048} integrates text features into the training of KT; and Guo, et al. \cite{bib049} trains the KT model by jointly training the original and adversarial samples.

\vspace{0.1cm}
\subsection{Affective  States Recognition Methods}

{\color{blue}Affect significantly influences learners' motivation and the learning state \cite{bib059,bib065}. Affective states recognition, a crucial aspect of affect computing, is increasingly accurate with Artificial Intelligence and sensor technology advancements. Educators commonly utilize affect scales to assess learners' affect, while physiological signals such as skin resistance and heart rate serve as reliable indicators \cite{bib050, bib017, bib051}.} 

In online learning, facial expressions and voice cues help understand students' engagement levels and KC mastery \cite{bib052, bib053, bib054}. Text conveys subtle affect nuances, complementing facial expressions and voice for comprehensive affect analysis \cite{bib055}. {\color{blue}Yang, et al. \cite{bib056} used deep learning techniques like BERT, LSTM, and ResNeST for effective multimodal affect recognition.} \textbf{In our DASKT model, affective states align closely with frustration, concentration, boredom, and confusion as outlined in \cite{bib001}}. 

{\color{blue}Behavior-affect topic modeling, seen in BSTM and TEAM, effectively integrates affect and behavior features for nuanced topic, affect, and behavior distributions \cite{bib057, bib058}. \textbf{In our DASKT model, affect simulation follows this joint modeling approach.}}

However, in our DASKT model, affect simulation is not entirely equivalent to behavior-affect topic modeling. Our model sources affect data from discrete records in the online tutoring system, unlike BSTM's rich interaction scenarios. Despite less rich affect properties, our data reflects students' independent learning objectively. Our dynamic affect simulation method ensures affect continuity, calculating affect for each problem, capturing spatiotemporal affect trajectories. Unlike the semester-based TEAM model, we model affect evolution at each timestep, integrating it with the student's knowledge states. This unified approach enhances model accuracy and offers insights into the learning process.

\section{Preliminary}

In this section, we provide formal definitions of KT and affect computation, involving affective factors. Additionally, we introduce some important embeddings involved in the DASKT framework proposed in this paper. Table \ref{table1} summarizes the mathematical symbols used in this paper.

\begin{table}[t]
\centering
\caption{\textbf{Mathematical symbol and descriptions}}
\label{table1}
\small
\begin{minipage}{0.5\textwidth}
\begin{tabularx}{\textwidth}{l>{\raggedright\arraybackslash}X}
\toprule
Symbol & Description \\
\midrule
S, P, R  & The set of student, problem and response\\
KC, Aff & The set of knowledge concept and the affect \\
n, j, k & The number of student, problem, and KC \\
$L$ & Students’ exercise sequence \\
$h_t$ & Students’ knowledge states at time t \\
$kc$, $\mathbf{KC}$ & The kc and its embedding \\
$p$, $\mathbf{P}$ & The problem and its embedding \\
$r$, $\mathbf{R}$ & Students’ response and its embedding \\
$Aff_j$, $\mathbf{Aff}$ & kc specific affect and its embedding \\
$L_{Seg}$  & Length of interval segment \\
MAF & Mining affective factors \\
ICA & Interval Computation of affect \\
DASE & Dynamic affect simulation effect \\
\bottomrule
\end{tabularx}
\end{minipage}
\end{table}

\subsection{Problem Definition}

In online intelligent tutoring systems, we define the set of students $S$, the set of problem IDs $P$, the set of knowledge concepts $KC$, the set of affect $Aff$, and the set of students' responses $R$. The set of students $S = \{s_1, s_2, \dots, s_n\}$ includes $n$ distinct students. The problem IDs set $P = \{p_1, p_2, \dots, p_j\}$ consists of $j$ distinct problem IDs. Each student can have $I (I \geq 1)$ records of problem responses, and the same problem can be responded to repeatedly. The knowledge concepts set $KC = \{kc_1, kc_2, \dots, kc_k\}$ contains $k$ distinct knowledge concepts, and each problem can be mapped to one or more corresponding knowledge concepts. Building on the research by Pardos, et al.\cite{bib001}, this study focuses on the primary affective  states in ASSISTments: \textbf{frustration, concentration, boredom, and confusion}. This paper adopts this states of affect, $Aff = \{aff_1, aff_2, aff_3, aff_4\}$ corresponding to the four different affect. The response set $R= \{0,1\}$ is a binary set used to mark whether the problem is answered correctly or not, where $0$ indicates an incorrect answer and $1$ indicates a correct answer.

Therefore, during the student's exercise process, we represent the student's response record as $L = \{(p_1, kc_1, aff_1, r_1), (p_2, kc_2, aff_2, r_2), \dots, (p_t, kc_t, aff_t, r_t)\}$ (in this paper, when a problem involves multiple knowledge concepts, we only take the first knowledge concept as the corresponding knowledge concept for the problem). Here, $p_t$ represents the problem answered by the student at time step $t$, $kc_t$ represents the knowledge concept of the corresponding problem, $aff_t$ represents the affect of the student at time step $t$, and $r_t$ represents the response label of the student to problem $p_t$ at time step $t$, i.e., the correctness of the answer. The length of such a exercise sequence is $t$. By analyzing the student's historical response records, we can track the student's mastery of knowledge concepts. The definition of knowledge tracking in this paper is as follows.

\noindent \textbf{Knowledge Tracing (KT)}: Given a student's exercise record $L = \{(p_1, kc_1, aff_1, r_1), \dots, (p_t, kc_t, aff_t, r_t)\}$, the KT task is a dynamic process. The student's mastery of knowledge changes over time, with the affect expressed during problem-solving, and the progression of exercise. The task aims to judge a student's mastery of KCs under the influence of affect based on their historical answering situations, and further predict the student's performance on the next KC. According to the different affect of students at different stages, corresponding feedback and intervention measures are provided. For example, for proactive students, encouragement and rewards can be given, and more advanced learning resources can be provided.

\subsection{Definition of Affective Factors}
Affect plays multiple roles in the learning process, influencing students' learning motivation, cognitive processes, learning experiences, and interactions. In educational practice, paying attention to affect education and students' affective  states has significant implications for promoting students' overall development and improving learning outcomes \cite{bib003,bib004}. {\color{blue}In light of this, this paper adopts the three affective factors proposed by \cite{bib002,bib061}, namely students' confidence, interest, and effort, for analysis.}  

\subsubsection{Students' Confidence}
Confidence can influence their perception of their own abilities, thereby shaping their learning attitudes and behaviors. This paper uses the accuracy  of KCs and the response time of KCs to simulate the calculation of student confidence. {\color{blue}In this study, we were inspired by Classical Test Theory (CTT) \cite{bib005} and relevant research by Sherbino, et al. \cite{bib062}, so we used prior statistical methods to evaluate the accuracy and response time of KCs.}  First, the accuracy  $A^i_k$ of KC for student ${s_i}$ and the response time $AT^i_k$ for each exercise corresponding to a certain time are defined as follows: 
\begin{equation}
    \begin{aligned}
        &A^{s_i}_k = \frac{1}{N^{s_i}_k} \sum_{n=1}^{C_k^{s_i}} \{r^{s_i}_{nk}==1\} ,
        \quad AT^{s_i}_k =   att^{s_i}_{tk}  , 	 
    \end{aligned}
\end{equation}
where $C_k^{s_i}$ represents the count of the $k$-th KC for student ${s_i}$, $r^{s_i}_{nk}$ is the label indicating whether the student ${s_i}$ answered correctly on the $k$-th KC in their own exercise sequence, and $N^{s_i}_k$ is the total occurrence of the $k$-th KC in the exercise sequence of student ${s_i}$. $att^{s_i}_{tk}$ is the time spent by student ${s_i}$ answering the $k$-th KC at the $t$-th time step. Then, the average accuracy  $\overline{A_k}$ and the average response time $\overline{AT_k}$ of the KCs for all students are defined as follows:
\begin{equation}
    \begin{aligned}
        &\overline{A_k} = \frac{1}{N_k} \sum_{i=1}^{|s_n|} A^{s_i}_k ,
        \quad \overline{AT_k} = \frac{1}{N_k} \sum_{i=1}^{|s_n|} AT^{s_i}_k  , 	 
    \end{aligned}
\end{equation}
where $|s_n|$ represents the size of the student set, and $N_k$ is the total occurrence of the $k$-th KC. Finally, the confidence of student ${s_i}$ is defined as:
\begin{equation}
    \begin{aligned}
        &Conf^{s_i}_k = concat(\sum_{k=1}^{|KC|} (A^{s_i}_k-\overline{A_k}), \sum_{k=1}^{|KC|} (AT^{s_i}_k- \overline{AT_k})),\\	 
    \end{aligned}
\end{equation}
where $|KC|$ represents the total number of KCs contained in the dataset. $concat(\cdot)$ is a concatenation function.

\subsubsection{Students' Interest}
Interest is an important source of learning motivation and it can enhance the cognitive process of learners. {\color{blue}Inspired by \cite{bib064}, this paper uses the length of students' exercise sequences to simulate interest and defines students' participation rate in the intelligent tutoring system based on these sequences.} The simulated students' participation rate remains consistent with the participation rate of the students before being segmented, which is defined as follows: 
\begin{equation}
    \begin{aligned}
        &PR^{s_i} = \frac{|el^{s_i}|}{t_e-t_s} ,\\	 
    \end{aligned}
\end{equation}
where $ |el^{s_i}|$ is the set of the exercise sequence for student ${s_i}$. $t_e$ and $t_s$ represent the start and end time of the student ${s_i}$'s exercise, respectively. The average participation rate of all students is defined as follows:
\begin{equation}
    \begin{aligned}
    &\overline{PR} = \frac{1}{|s_n|}\sum_{i=1}^{|s_n|} PR^{s_i},\\	 
    \end{aligned}
\end{equation}
where $|s_n|$ is the set of all students. Therefore, the interest of student ${s_i}$ is defined as follows:
\begin{equation}
    \begin{aligned}
        &Inter^{s_i} =PR^{s_i} - \overline{PR} .\\	 
    \end{aligned}
\end{equation}

\begin{figure*}
    \centering
    \includegraphics[width=1.0\textwidth]{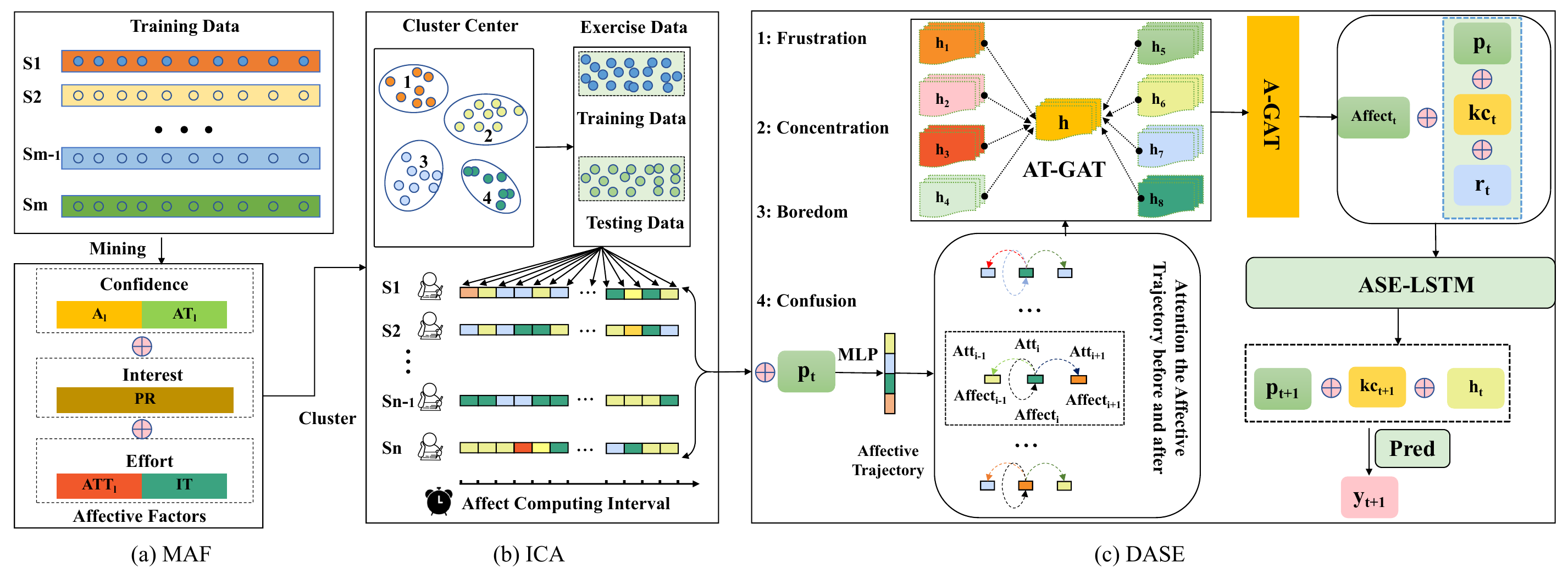}
    \caption{\textbf{The main structure of our DASKT model is presented. The diagram illustrates the method of obtaining affective factors and the process of calculating affect at intervals. Based on this, at time step $t$, the inputs include affect $affect_t$, problem embedding $P_t$, knowledge concept of the problem $KC_t$, student's response $r_t$, and knowledge states $h_t$. In addition, we can also predict the student's performance at time step $t+1$, denoted as $y_{t+1}$.}\label{figure02}}
\end{figure*}

\subsubsection{Students' Effort}
Effort directly impacts their learning performance. Through diligent study, students cultivate the ability to solve problems and overcome difficulties. {\color{blue}Inspired by \cite{bib063}, we use the number of attempts and the interval between questions for different KCs as indicators to calculate effort.} The number of attempts at KCs at different time steps $Att^{s_i}_k$ and the average number of attempts at this KC by all students $\overline{Att_k}$ are defined as follows: 
\begin{equation}
    \begin{aligned}
        &Att^{s_i}_k =  \sum_{t=1}^{|el^{s_i}|} att^t_k ,
        \quad \overline{Att_k} =  \frac{1}{|s_n|} \sum_{i=1}^{|s_n|} Att^{s_i}_k , 	\\
    \end{aligned}
\end{equation}
where $att^t_k$ is the number of attempts by student ${s_i}$ at KC of $k$ at time step $t$.
The interval between the previous problem and the next problem in the exercise sequence of student ${s_i}$, $IT^{s_i}$, and the average answer time interval of all students, $\overline{IT}$, are defined as follows:
\begin{equation}
    \begin{aligned}
        &IT^{s_i} = \sum_{t=1}^{|el^{s_i}|} p_t-p_{t-1},
        \quad \overline{IT} = \frac{1}{|s_n|}\sum_{i=1}^{|s_n|} IT^{s_i},
    \end{aligned}
\end{equation}
where $p_t$ represents the starting time of the problem at time $t$ for student $s_i$, and $p_{t-1}$ denotes the starting time of the problem at time $t-1$ for the same student $s_i$. Consequently, the effort exerted by student $s_i$ can be defined as follows:
\begin{equation}
    \begin{aligned}
        &Eff^{s_i} = concat(\sum_{k=1}^{|KC|} (Att^{s_i}_k-\overline{Att_k}), IT^{s_i}- \overline{IT}).\\	 
    \end{aligned}
\end{equation}

\subsection{Embeddings}
Before delving into the details of the DASKT structure, we can first provide a brief overview of the basic embeddings in DASKT, which will help better understand how DASKT operates internally.

\subsubsection{Problem IDs, KCs and Responses Embeddings}

In the DASKT model, the basic elements include Problem ID, KC, and Student Response. Specifically, we use an embedding matrix \( \mathbf{P} \in \mathbb{R}^{j \times d_p} \) (where \( d_p \) is the embedding size and j is the number of all different problems) to represent the IDs of all problems answered by students. On the other hand, we use an embedding matrix \( \mathbf{KC} \in \mathbb{R}^{k \times d_k} \) (where \( d_k \) is the embedding size and k is the number of all different knowledge concepts) to represent all knowledge concepts associated with the problems. In addition, we use \( \mathbf{R} \in \mathbb{R}^{2 \times d_r} \) (where \( d_r \) is the embedding size and 2 is the number of types of answers, with 1 representing a correct answer to the problem and 0 representing a wrong answer) to represent the responses of students.

\subsubsection{Affect Embedding}
Affect is an important variable in the model proposed in this paper, and its embedding matrix is represented as $\mathbf{Aff} \in \mathbb{R}^{4 \times d_{aff}}$ ($d_{aff}$ is the embedding size and 4 is the number of states of affect). Affect is summarized into these four states based on the phenomena often exhibited by students in online intelligent tutoring systems: frustration, concentration, boredom, and confusion.

\section{The DASKT Framework}
In this section, we will provide a detailed introduction to the DASKT framework proposed in this paper. The framework consists of three parts, as shown in Figure \ref{figure02}. First, we mine the training set to count the affective factors such as confidence, interest, and effort in affective factors, initially forming the mining affective factors module (MAF), as shown in Figure \ref{figure02} (a). We then perform cluster learning on the affective factors and interval-based affect calculation (ICA) for the training and test sets based on the cluster centers, as shown in Figure \ref{figure02} (b). Finally, we use the graph attention network for dynamic affect simulation and apply the learned  affect to the KT task under the effect of affect  (DASE), as shown in Figure \ref{figure02} (c). Subsequently, we will provide a more detailed introduction to these three modules.

\subsection{MAF: Mining Affective Factors}

Different affect of students towards learning directly affect their learning outcomes. Positive affect can stimulate learning motivation, regulate cognitive processes, improve memory levels, and deepen understanding of knowledge \cite{bib006,bib007}. Negative affect, on the other hand, can reduce learners' participation and interest in learning, influence learners' reprocessing of knowledge, and result in learning outcomes below the normal level \cite{bib008,bib009}. However, the current understanding of students' affect mainly relies on the use of scale measurements, physiological signals, facial expressions, and voice signals for assessment. These judgment methods not only require a lot of time but also require a large number of hardware devices to collect students' affective factors. Especially, in online learning, the inability of teachers to observe students' performance face-to-face makes this phenomenon more prominent. Therefore, it is very meaningful to design a way that can simulate the calculation of students' real affect through students' online historical response data. This paper proposes an MAF module, which is responsible for mining some necessary affective factors needed to simulate student affective  states. The specific calculation method is as follows:

\begin{equation}
    \begin{aligned}
        &AF^{s_i} = concat(Conf^{s_i}_k,Eff^{s_i},Inter^{s_i}),\\
        &\hat{AF^{s_i}} = \frac{AF^{s_i}}{\| {AF^{s_i}} \|_2}, 
        \quad \hat{AF} = \hat{AF^{s_i}}_{i=1}^{|s_m|},
    \end{aligned}
\end{equation}
where $Conf^{s_i}_k$, $Inter^{s_i}$, and $Eff^{s_i}$ are confidence, interest, and effort in the affective factors of the  student $s_i$, respectively. $\hat{AF^{s_i}}$ is the $L2$ normalization of affective factors. $|s_m|$ is the set of all students in the training set, and $\hat{AF}$ is the summary of affective factors of $|s_m|$.

\subsection{ICA: Interval Computation of Affect}  
After obtaining the affective factors $\hat{AF}$, it is necessary to identify the affective  states for different students. This paper uses clustering to find the clustering centers of affective  states in the training set, in order to calculate the four different affect clustering centers. {\color{blue}The K-means algorithm \cite{bib010} is used in this paper to calculate the affect clustering centers.} $Aff^r_{c}$ is the randomly selected affect clustering center, $\hat{AF}_c$ is the clustering set where all student affect in the training set are located, and $Aff_{c}$ is the new clustering center set. The calculation process is as follows:
\begin{equation}
    \begin{aligned}
        &Aff^r_{c} = rand(\hat{AF}[:|C|]),\\
        &\hat{AF}_c = [\text{argmin}_{j=1}^{|C|} (\hat{AF}  - {Aff^j_{c}})^2 ],\\
        &Aff_{c} =[  \frac{1}{n} \sum_{i}^{|C|}\hat{AF}^i_c ],
    \end{aligned}
\end{equation}
where $|C|$ is the number of affect clusters, \( \text{rand}(\cdot) \) is a random function, \( \text{argmin}(\cdot) \) is a minimum function. \( \text{Aff}^j_{c} \) is the \(j\)-th cluster center. \( n \) represents the number of data in this affective  state, and \( \hat{AF}^i_c \) represents the sum of all data  in the \(i\)-th affective  state.

\begin{figure} 
    \centering
    \includegraphics[width=0.5\textwidth]{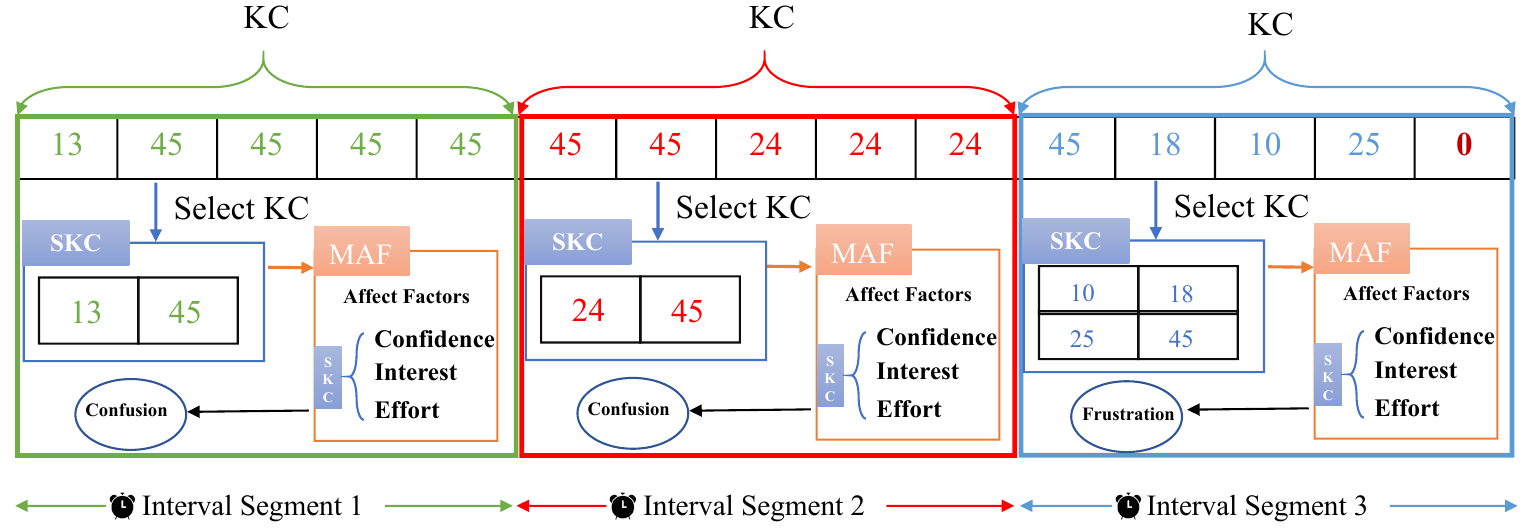}
    \caption{\textbf{An illustrative example of interval-based affect calculation is shown where a student's sequence of 14 attempts is split into three interval segments, correlating to time intervals for answering 5 problems each. If the student ceases interaction, remaining segments are zero-filled. Attempt counts vary according to the number of problems addressed.}\label{figure03}}
\end{figure}

The affective  states of students cannot be constant throughout the exercise sequence process. Different stages show different affective  states, which is more in line with the real affect expression of students. {\color{blue}Inspired by Minn, et al.\cite{bib011} on the dynamic interval assessment of student abilities, in order to evaluate the affective  states of students in stages, we adopted a method based on time intervals.} The time interval, denoted as $L_{Seg}$, is defined as a segment that contains the number of times a student answers questions within the system. Before the next time interval begins, based on the evaluation of the student's previous performance history, the student's affect are evaluated for each time interval through clustering, as shown in figure \ref{figure03}. 
Because many KCs are included in the student's exercise sequence, after each KC is processed by the MAF  module, the vector size becomes three times the original, which will consume a lot of computational memory. By dividing each student's attempt sequence into multiple time intervals and only evaluating the affect situation of the KCs in the segment each time, this can reduce the computational burden and optimize memory space allocation in order to more effectively learn the entire long sequence, resulting in $\hat{AF^{s_i}_{seg}}$. Furthermore, after each time interval, the student's affective  states $\hat{AF^{seg}_c}$ is reassessed to prepare for the next step of dynamically assessing the student's affect. The specific calculation method is as follows:
\begin{equation}
    \begin{aligned}
        &  N = \frac{\hat{AF^{s_i}}}{L_{Seg}}, \\
       &  \hat{AF^{s_i}_{seg}} = \{MAF(\hat{AF^{s_i}_1}), MAF(\hat{AF^{s_i}_2}), \ldots, MAF(\hat{AF^{s_i}_N})\} ,\\
        &\hat{AF^{seg}_c} = ((\text{argmin}_{j=1}^{|C|} (\hat{AF^{s_i}_{seg}}  - {Aff^{j}_{c}})^2)_{seg=1}^{N})_{s_i=1}^{|s_n|} ,\\
    \end{aligned}
\end{equation}
where $L_{Seg}$ is the length of the interval segment, $N$ is the number of affective  states of student $s_i$. $MAF(\cdot)$ is the Mining Affective Factors function module. $|s_n|$ is the set of all students in the training and test sets.

\subsection{DASE: Dynamic Affect Simulation Effect} 
{\color{blue}The affective  states of students have been assessed intermittently through the ICA module. However, through relevant psychological research, it is found that changes in affect are always fluctuating and coherent, and there is a strong correlation between recent affective  states \cite{bib013,bib014,bib015}.} Although intermittent affect assessment has to some extent simulated the fluctuations of student affect, how to effectively simulate the coherence of student affect still needs to be further resolved. This paper proposes the dynamic affect simulation module, which can effectively simulate the continuity of human affect changes. Specifically, we first bind the affect evaluated by the ICA module with the corresponding problems in the learning exercise process, dynamically learning the changes in student affect as the exercise deepens. The preliminary calculation of the affect trajectory $Atra^{s_i}_t$ of student $s_i$ at time $t$ is as follows:
\begin{equation}
    \begin{aligned}
        &  Atra^{s_i}_t =  W_1(\hat{AF^{s_i}_t} \oplus  p_t)+b_1, \quad \hat{AF^{s_i}_t} \in \hat{AF^{seg}_c},\\
    \end{aligned}
\end{equation}
where $\oplus $ is vector concatenation, $W_1 \in \mathbb{R}^{ (d_{aff}) \times  (d_{aff} + d_{p})}$ is a learnable weight parameter, $b_1 \in \mathbb{R}^{ d_{aff}} $ is a bias term, and $\hat{AF^{s_i}_t}$ is the affect of student $s_i$ at time $t$.

Then we construct the affect trajectory graph. Since there is no well-constructed relationship between affect nodes and edges in the dataset, this paper uses the affect at time $t$ and the affect at times $t-1$ and $t+1$ to construct the nodes and edges of the affect graph. That is, it is believed that the current affect has a deep coherence with the affect at the previous and next moments and should be greatly influenced by the previous and next affect. The construction of the affect graph of student $s_i$ is as follows:
\begin{equation}
   \footnotesize
    \begin{aligned}
        & V^{s_i}_{aff} = \{ Atra^{s_i}_1,Atra^{s_i}_2, \ldots, Atra^{s_i}_{N}\} \mid |V_{aff}| = N ,\\
        & E^{s_i}_{aff}=  \{ (Atra^{s_i}_j, Atra^{s_i}_{j-1}), (Atra^{s_i}_j, Atra^{s_i}_{j}), (Atra^{s_i}_j, Atra^{s_i}_{j+1})  \}_{j=1}^{N},\\
        & G^{s_i}_{aff} =(V^{s_i}_{aff},E^{s_i}_{aff}),
    \end{aligned}
\end{equation}
where $V^{s_i}_{aff}$ and $E^{s_i}_{aff}$ represent the collection of affect nodes and affect edges of student $s_i$, respectively. $|V_{aff}|$ is the size of the affect node set.

After the affect graph is constructed, it is necessary to learn the coherence features of the affective trajectory based on this graph. The aggregation and update operations in GAT \cite{bib016} are used to represent affect coherence features, $AT\text{-}GAT_n^{(1)}$ is the affect trajectory representation between the $n$-th affect and the $j$-th affect of student $s_i$ using the $AT\text{-}GAT$ network. $A\text{-}GAT_n^{(2)}$ is the dynamic affect representation between the $n$-th affect and the $j$-th affect of student $s_i$ using the $A\text{-}GAT$ network, the specific operations are as follows:
\begin{equation}
\footnotesize
    \begin{aligned}
        &AT\text{-}GAT_n^{(1)} = ELU \left( \frac{1}{|h|}\text{concat}(\{ \sum_{j \in {E^{s_i}_{aff}}(n)} \alpha_{nj}^{(k)} W_k^{(1)} {V^{s_i}_{aff}}_j \}_{k=1}^{|h|}) \right),\\
        &A\text{-}GAT_n^{(2)} = ELU \left( \sum_{j \in {E^{s_i}_{aff}}(n)} \alpha_{nj} W^{(2)} {AT\text{-}GAT_n^{(1)}}_j \right),
    \end{aligned}
\end{equation}
where $ELU$ is a nonlinear activation function. $|h|$ is the collection of multi-head attention mechanisms, $\alpha_{nj}^{(k)}$ is the $k$-th attention coefficient between the $n$-th affect and its $j$-th neighbor affect. $W_k^{(1)}, W^{(2)}\in \mathbb{R}^{ (d_{aff}) \times  (d_{aff})}$ are trainable weight parameters.

In order to explore the impact of affect on students' knowledge states, we use affect-enhanced problem feature representation. In view of this, we jointly learn the generated dynamic affect \( A\text{-}GAT_n^{(2)} \) with the problems \( p_t \), KC  of \( kc_t \), and student responses \( r_t \) experienced in the student learning sequence, thereby obtaining the knowledge tracking under the effect of the continuously developing affect of the student. Specifically, we use the information fusion affect method to enrich the student's learning representation, and then use an affect simulation effect layer (ASE-LSTM) to judge the student's knowledge status under the current affect. The formula is as follows:
\begin{equation}
    \begin{aligned}
        &h_{t}=LSTM (p_t \oplus kc_t \oplus A\text{-}GAT_n^{(2)} \oplus r_t) , \\
    \end{aligned}
\end{equation}
where LSTM stands for Long Short-Term Memory, $h_{t} \in \mathbb{R}^{d \times d}$ is the manifestation of the student's knowledge status under the affect utility at time $t$.

\subsection{Prediction and Objective Function}
Since we have obtained the knowledge status \( h_t \) of the student at the current time step \( t \) under the affective  states through the \text{DASE} network, we fuse the problem \( p_{t+1} \) and KC of \( kc_{t+1} \) at time step \( t+1 \) to enhance the representation of the student's application of learned knowledge under the affect boost. The formula is as follows:
\begin{equation}
    \begin{aligned}
        &y_{t}=p_{t+1} \oplus kc_{t+1} \oplus h_t , \\
        &y_{t+1}=\sigma (w_t  y_{t} +b_t),
    \end{aligned}
\end{equation}
where \( w_t \in \mathbb{R}^{1 \times (d_p+d_k+d) } \) is a learnable weight parameter, \( b^1_t \in \mathbb{R}^{1} \) is a bias term, and \( \sigma (\cdot) \) is a sigmoid nonlinear activation function.

To train all the parameters in the DASKT  framework, we use the cross-entropy log loss between the framework prediction value \( y \) and the correctness of the student's answer to the question \( r \) as the objective function:
\begin{equation}
    \begin{aligned}
        &\mathcal{L} (\theta) = -\sum_{t=1}^T  (r_t \log (y_t) +  (1 - r_t) \log (1 - y_t)) + \lambda_\theta ||\theta||^2,\\
    \end{aligned}
\end{equation}
where \( \theta \) represents all the parameters of the DASKT framework, and \( \lambda_\theta \) represents the regularization hyperparameter. The Adam optimizer is used to minimize the objective function. The experimental part will elaborate on the details of other settings more detailedly.

\section{EXPERIMENTS}
In this section, we first introduce the real datasets used in the experiment, then detail the training process and baseline models. Then, we compare the results of the baseline models and the model proposed in this paper in predicting students' future performance. In addition, to better display the DASKT model, we answer the following research questions:
\begin{itemize}
\item \textbf{RQ1} Does the affect simulation calculation method we propose have accuracy and reliability?
\item \textbf{RQ2} Does the DASKT model we propose have good performance in predicting students' performance?
\item \textbf{RQ3} Does the DASKT model we propose maintain the consistency of the students' constantly changing knowledge states and affect journey?
\item \textbf{RQ4} How do the affective factors, interval affect computation, and dynamic affect  state in DASKT affect the knowledge tracing results?
\item \textbf{RQ5} How does  affect impact the performance of the DASKT model?
\end{itemize}

\subsection{Datasets}
For the calculation of affect simulation, this paper needs students' behavior such as problem IDs, KCs, answer time, interval time between answers, the number of attempts for each exercise, and other related attributes that can be calculated based on the answer situation. {\color{blue}To our knowledge, only the real-world datasets ASSIST2012 and ASSISTchall meet these requirements, and no real affect labels were used in training. }The statistical data of the datasets are shown in Table \ref{table2}. {\color{blue}In this study, the exercise sequence lengths are standardized to 100 for the ASSIST2012 dataset and 500 for the ASSISTchall dataset. Sequences longer than these thresholds will be divided into segments representing virtual students, while shorter sequences will be padded with zeros to meet the required length. Detailed descriptions of the datasets are as follows:} 
\begin{table}[h]
    \centering
    \caption{\textbf{Statistics of all datasets.}}
    \resizebox{0.46\textwidth}{!}{%
        \begin{tabular}{lccc}
            \toprule
            \diagbox{Statistics}{Datasets} & ASSIST2012 & ASSISTchall  \\
            \midrule
            Number of exercise records & 2711813 & 942816  \\
            Number of students & 29018 & 1709  \\
            Number of problem IDs & 53091 & 3162  \\
            Number of KCs & 265 & 102  \\
            Number of answer time & 2711813 & 918115  \\
            Number of interval time & 2711813 & 942816  \\
            Number of attempt counts  &2700324 &942816  \\   
            Number of records with affect  & 2711813 & 918115  \\
            \bottomrule
        \end{tabular}
    }
    \label{table2}
\end{table}

\begin{itemize}
   \item ASSIST2012\footnote{https://sites.google.com/site/assistmentsdata/home/2012-13-school-data-with-affect} is an innovative online tutoring system that has been operational since 2004. The comprehensive dataset originates from skill-building problem sets, meticulously designed for students to complete similar exercises until mastery is achieved. It includes detailed affect prediction data for the academic year 2012-2013. Records lacking essential knowledge concepts have been systematically filtered out from this dataset.
   \item ASSISTchall\footnote{https://sites.google.com/view/assistmentsdatamining/dataset} derived from a comprehensive longitudinal study conducted from 2004 to 2007, meticulously tracks students' interactions with the ASSISTments blended learning platform throughout their middle school journey. Distinguished by the notably extended learning trajectories of its students, especially when compared to those in the ASSIST2012 dataset, ASSISTchall presents a uniquely challenging dataset crafted to evaluate students' learning abilities.  It is specifically designed to delve into students' capabilities and problem-solving skills across a diverse range of knowledge domains. It also includes detailed affect prediction data. 
\end{itemize}

\subsection{Experimental Setup}

{\color{blue}In our experiments, we filtered out records with empty "problem ID" or "KC" fields and sorted the remaining data accordingly. For the ASSISTchall dataset, sorting was based on the "startTime" field, while for the ASSIST2012 dataset, sorting was based on the "start\_time" field. All model parameters were randomly initialized during training. For all datasets, we performed standard 5-fold cross-validation for all models. In each fold, 80\% of the student data was split into a training set (80\%) and a validation set (20\%), with the remaining 20\% used as the test set. To ensure fairness, we fine-tuned all comparison algorithms to achieve optimal performance on our experimental datasets. All hyperparameters were learned on the training set, and the model that performed best on the validation set was used to evaluate the test set. In our DASKT implementation, we maintained consistent dimensional parameters (e.g., dp, dk, dr, daff, and d), all set to 256 to streamline the model design. A batch size of 32 was used, and the experiments were conducted on a 64-bit Ubuntu 20.04.5 LTS server equipped with an Intel Xeon Gold 5218 CPU @ 2.30GHz and a single 32GB Tesla-V100 GPU, running on the PyTorch framework.} 

\subsection{Benchmark Methods}

{\color{blue} To evaluate the effectiveness of DASKT, we compare it against 12 different baseline methods. All baseline methods use the same training, validation, and test sets as this paper. In our experiments, we strictly adhere to the recommended parameters from the original papers or publicly available implementations of each algorithm. To ensure fairness, we fine-tune the parameters of all comparison algorithms to achieve their optimal performance on our experimental datasets. Our comparative analysis is based on four performance metrics: Root Mean Square Error (RMSE), Area Under the Curve (AUC), Accuracy (ACC), and the Square of Pearson Correlation (r²).}

\begin{itemize}
  \item  \textbf{DKT}\cite{bib019} is the pioneering KT model that integrates deep learning and depends on Recurrent Neural Networks (RNN) or Long Short-Term Memory (LSTM) to infer students' knowledge proficiency.
    \item \textbf{DKT-DSC}\cite{bib011} employs K-means clustering to group students with similar skill levels into clusters regularly. By assigning each student to the nearest cluster based on centroid proximity, this method aids in predictive modeling.
    \item \textbf{DKVMN}\cite{bib020} enhances memory-augmented neural networks with a key-value memory structure. Keys access relevant memory slots while values consolidate data, offering a richer and adaptable representation of student knowledge.
    \item \textbf{SAKT}\cite{bib021} model, by utilizing a self-attention mechanism, evaluates the significance of every past interaction. Consequently, the model is capable of concentrating on the most pertinent past interactions during predictions, rather than treating all historical interactions uniformly.
   \item  \textbf{CKT}\cite{bib022}  model integrates hierarchical convolutional layers to derive adaptive learning rates from students' continuous learning interactions. This innovative approach facilitates the automatic acquisition of meaningful exercise embeddings.
   \item \textbf{GKT} \cite{bib029} pioneers the use of Graph Neural Networks (GNN) in knowledge tracing. By converting the knowledge framework into a graph, it reframes knowledge tracing as a time-series node-level classification challenge within GNN. To address the ambiguity of knowledge graph structures, the model presents diverse graph structure implementations.
    \item \textbf{EKT} \cite{bib023} considers the significance of each exercise with respect to all knowledge concepts, utilizing the semantic details of the exercises as input.
    \item \textbf{JKT} \cite{bib024} employs graph structures to capture intricate relationships between exercises and concepts, boosting interpretive capabilities by connecting exercises across multiple concepts and revealing nuanced semantic insights.
    \item \textbf{LPKT}  \cite{bib025} tracks students' knowledge states by directly simulating their learning process. It incorporates a Learning Gain module to mimic students' capacity to acquire knowledge and a Forgetting Effect module to replicate the decay of knowledge retention over time.
    \item \textbf{DIMKT} \cite{bib026} incorporates difficulty hierarchies into problem representation. It introduces a three-stage adaptive sequential neural network to correlate a student's knowledge status with problem difficulty levels during practice sessions.
    \item  \textbf{QIKT}  \cite{bib027} explicitly captures the nuances of a student's knowledge states evolution by acquiring problem-specific cognitive representations through a knowledge acquisition module centered around problems and a question-solving module.
    \item  \textbf{TCKT} \cite{bib028} introduces a causal self-attention mechanism grounded in the front-door adjustment theory, enhancing interactive representation and mitigating prediction errors stemming from dataset biases.
\end{itemize}

\subsection{Affect Simulation Calculation: Accuracy and Reliability (RQ1)}

{\color{blue}Affect detection is complex due to physiological influences and the lack of specialized hardware \cite{bib017,bib018}. The affect detector \cite{bib001}, developed using log files and field observations from the ASSISTments system, generates confidence levels for four affect types—frustration, concentration, boredom, and confusion—on the ASSIST2012 dataset. For comparison with the DASE module, we use the affect with the highest confidence for each exercise, as shown in Table \ref{table3}.} 

\begin{table}[h]
\caption{\color{blue}{\textbf{Comparison (\%) of affect simulation effects. "Consistency Rate" refers to a comparison between the affect detected by the affect detector (AD) for each question and the consistency of affect judged by the proposed DASE module in this paper. The affect determined by these two approaches are then applied to the DKT model separately for a comparison of the AUC effects.}}}
\renewcommand{\arraystretch}{2.0} 
\Huge  
\resizebox{0.48\textwidth}{!}{%
\begin{tabular}{cccccc}
\toprule
Affect & Consistency Rate  &  DKT-AD & DKT-DASE  & GKT-AD & GKT-DASE \\
\midrule
Frustration & 78.19 & 15.27 & 16.22 & 15.89 & 20.13 \\
Concentration & 74.57 & 20.17 & 23.08 & 25.28 & 19.26 \\
Boredom & 80.29 & 16.35 & 25.29 & 24.55 & 18.42 \\    
Confusion & 72.74 & 21.37 & 10.65 & 8.67 & 17.77 \\
Total & - & \textbf{73.16$^{ \uparrow0.12}$} & \textbf{75.24$^{ \uparrow2.20}$} & \textbf{74.39$^{ \uparrow0.89}$} & \textbf{75.58$^{ \uparrow2.08}$} \\
\bottomrule
\end{tabular}}
\label{table3}
\end{table}

{\color{blue}The experimental results show that the affect predictions made by the affect detector on the ASSIST2012 dataset exhibit a high level of consistency with the DASE affect simulation calculation scheme proposed in this paper. Specifically, the consistency for detecting boredom reached 80.29\%. This high consistency may be attributed to the more distinct characteristics and manifestations of boredom compared to other affects, making it easier to detect. When applying the affect detector to the DKT and GKT models, both models showed performance improvements, with AUC increases of 0.12\% and 0.89\% compared to the original DKT and GKT models, respectively. Furthermore, when the proposed DASE affect simulation method was incorporated, the AUC performances of the DKT-DASE and GKT-DASE models improved by 2.20\% and 2.08\%, respectively. These findings further validate the high accuracy and reliability of the DASE affect calculation method and highlight the crucial role of accurately capturing affective factors in enhancing knowledge tracing models.} 

\subsection{Future Performance Prediction (RQ2)}

\begin{table*}
\centering
\caption{\textbf{On the ASSIST2012 and ASSISTchall datasets, a comparison of all methods was conducted for the task of predicting student performance based on four performance metrics. The state-of-the-art results were underscored, while the best results were highlighted in bold. In the outcomes, * denotes a p-value less than 0.05 in t-test.}}
\resizebox{\textwidth}{!}{%
\begin{tabular}{c|c|c|c|c|c|c|c|c|c|c|c|c|c|c|c}
\toprule
Datasets & Metrics & DKT & DKVMN & DKT-DSC & SAKT & EKT & GKT & CKT & JKT & LPKT & DIMKT & QIKT & TCKT &\textbf{DASKT (Ours)}   \\
\midrule
\multirow{4}{*}{ASSIST2012} & RMSE & 0.4226 & 0.4243 & 0.4268 & 0.4235 & 0.4070 & 0.4238 & 0.4213 & 0.4059 & 0.4037 & \underline{0.4006} & 0.4221 & 0.4017& \textbf{0.3849*}  \\
\arrayrulecolor{lightgray} 
\cmidrule(lr){2-15}
& ACC & 0.7375 & 0.7353 & 0.7266 & 0.7364 & 0.7594 & 0.7143 & 0.7391 & 0.7429 & 0.7634 & 0.7666 &0.7415 & \underline{0.7742} & \textbf{0.7813*}  \\
\arrayrulecolor{lightgray} 
\cmidrule(lr){2-15}
& AUC & 0.7304 & 0.7255 & 0.7210 & 0.7279 & 0.7732 & 0.7350 & 0.7351 & 0.7635 & 0.7824 & \underline{0.7899} & 0.7492 & 0.7763& \textbf{0.7925*}  \\
\arrayrulecolor{lightgray} 
\cmidrule(lr){2-15}
& $r^2$ & 0.1468 & 0.1398 & 0.0754 & 0.1403 & 0.1574 & 0.1462 & 0.1497 & 0.2105 & 0.2145 & \underline{0.2187 }& 0.2116 &0.2048 & \textbf{0.2248*}  \\
\arrayrulecolor{black} 
\midrule
\multirow{4}{*}{ASSISTchall} & RMSE & 0.4298 & 0.4326 & 0.6953 & 0.4311 & 0.4137 & 0.4151 & 0.4302 & 0.4218 & 0.4133 &\underline{0.4113}  & 0.4251 &0.4196  & \textbf{0.3926*}  \\
\arrayrulecolor{lightgray} 
\cmidrule(lr){2-15}
& ACC & 0.7202 & 0.7157 & 0.4281 & 0.7180 & 0.7435 &\underline{0.7541}  & 0.7193 & 0.7370 & 0.7446 & 0.7471 &0.7259 &0.7351 & \textbf{0.8215*}  \\
\arrayrulecolor{lightgray} 
\cmidrule(lr){2-15}
& AUC & 0.7649 & 0.7576 & 0.430 & 0.7620 & 0.8024 & 0.8158 & 0.7644 & 0.7450 & 0.8042 & \underline{0.8074} & 0.7428 &0.7965 & \textbf{0.8529*}  \\
\arrayrulecolor{lightgray} 
\cmidrule(lr){2-15}
& $r^2$ & 0.1425 & 0.1302 & 0.0984 & 0.0822 & 0.2042 & 0.2011 & 0.1488 & 0.1826 & 0.2605 & \underline{0.2611}  & 0.1976 & 0.3049 & \textbf{0.3246*}  \\
\arrayrulecolor{black} 
\bottomrule
\end{tabular}%
}
\label{table4}
\end{table*}

The core goal of the DASKT model is to establish a reasonable knowledge states model based on students' learning history, with a further aim to provide effective motivation or intervention for students in different affective  states. Although our goal is not simply to predict students' future performance, as the accuracy of predicting students' future performance increases, our estimation of their knowledge states also becomes more precise. To assess the ability of the proposed DASKT model to predict students' future performance, we compared this model with the aforementioned benchmark models. Specifically, for each student's sequence of exercises, except for the student's response to the first question which does not participate in future prediction, the responses to all other questions participate in the prediction at the corresponding time points. Through this comparison, we can better understand the potential and advantages of the DASKT model in the field of education, thereby providing more effective support for personalized learning and educational intervention.

To ensure the robustness of the evaluation results, we used four evaluation metrics in all experiments, namely Root Mean Square Error (RMSE), Accuracy (ACC), Area Under the Curve (AUC), and the square of the Pearson correlation coefficient ($r^2$). When calculating accuracy, we set the threshold to 0.5. Table \ref{table4} presents the detailed experimental results, and also shows the statistical significance of our model compared to the best baseline model. From Table \ref{table4}, we made several important observations. First, DASKT is significantly superior to all baseline methods on all datasets and evaluation indicators. This indicates that considering the affective factors in the student's answering process on student learning is necessary and valuable. Second, in the ASSISTchall dataset, DASKT is significantly superior to the most advanced DIMKT model (i.e., AUC increased by 4.55\%), which shows that DASKT is more capable of capturing students' historical learning information with long sequences. In addition, in the ASSIST2012 dataset with affect, DASKT is also slightly superior to other methods. Third, the existing best KT models that use graph neural networks to construct relationships between KCs (such as GKT and JKT) performed well, further proving the applicability of graph networks in the KT field. Fourth, compared to other knowledge tracing methods that do not consider affective factors, DASKT further considers affective  states in practice, successfully evaluating students' affective  states in different exercises, establishing the relationship between student knowledge states and affect, and achieving better results. Finally, compared with LPKT, which directly simulates the student learning process, DASKT performed better, which encourages us to consider affect more deeply in the learning process and lays the foundation for greater success in the future.These findings will open up a new avenue for research in the field of knowledge tracing and offer insightful guidance for advancing both research and practical applications.

\begin{figure}[!htb]
    \centering
    \includegraphics[width=0.5\textwidth]{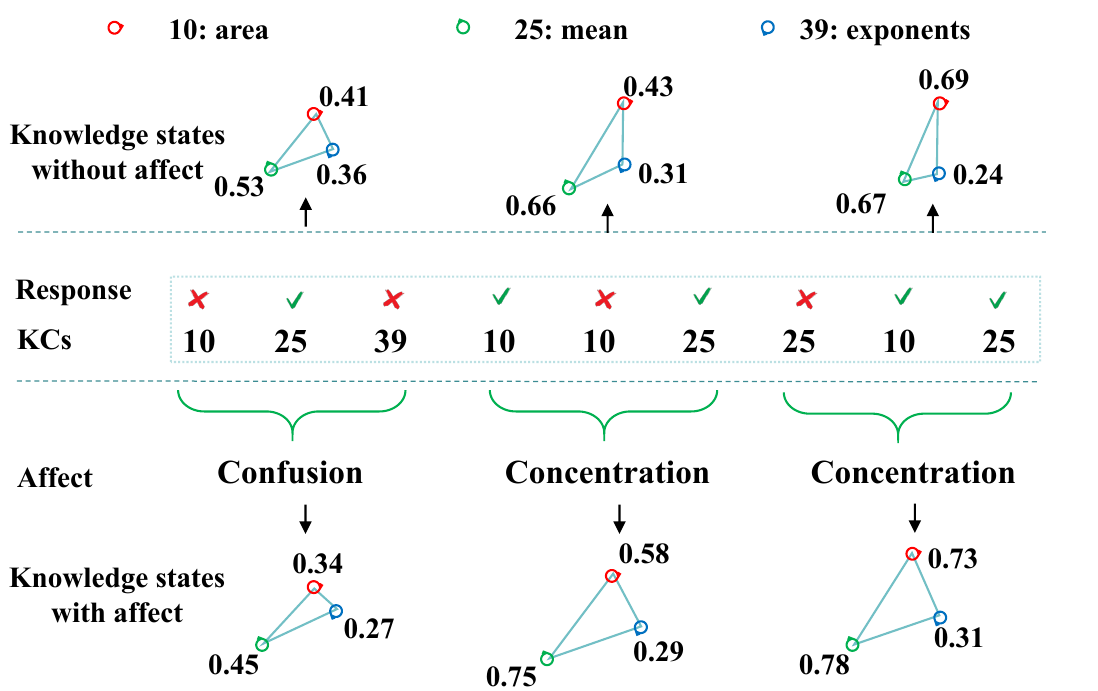}
    \caption{\textbf{This is a comparative diagram that illustrates the differences in a student's knowledge states when completing a sequence of nine  exercises, with and without affect involvement. The knowledge states is updated every three exercises, with  numerals representing the value of the knowledge states. A higher value indicates a better predicted mastery of the corresponding KC. The top half of the diagram showcases the knowledge states in the absence of affect, while the bottom half depicts the knowledge states when affect are factored in.}\label{figure06}}
\end{figure}

\subsection{Analysis of the Consistency between knowledge states and affective  states (RQ3)}
{\color{blue} Given our primary objective of modeling the learning process of students, we will demonstrate that DASKT can effectively capture students' knowledge states in the context of affective involvement and align with their affective trajectories. Figure \ref{figure06} shows the changes in the knowledge states tracked by DASKT when a student completes part of the exercise records extracted from the same exercise sequence. There are several key observations in the figure.

Firstly, our proposed DASKT method can capture different knowledge states that students acquire in learning interactions under affective  states. For example, when introducing affective  states, it can be clearly seen that students' knowledge states in the same exercise sequence are different, which proves the important role of affect in the process of knowledge states modeling. 

Secondly, we found that affect have a significant impact on the state of knowledge. Taking the KC of 39  (exponents) as an example, this KC only appears once in the entire exercise sequence, and the student's answer is incorrect. Without considering affect, the model will gradually recognize this knowledge state as a decrease in mastery. However, when students are in an affective state dominated by concentration, there is actually a slight improvement in their mastery of the KC.  Moreover, upon a meticulous examination of the knowledge states across three KCs, it is observable that under the influence of positive affective states, the overall enhancement in knowledge states tends to be marginally superior compared to scenarios where affective factors are not taken into account, even with identical feedback from exercises. Conversely, under the influence of negative affective states, the overall decline in knowledge states is generally slightly more pronounced than in cases where affective factors are disregarded.  This discovery is consistent with the view that positive affect can overall enhance students' memory and understanding, thereby enhancing their understanding of KCs.

Lastly, we observe that within the same exercise sequence, positive affect significantly enhances the mastery and retention of  KCs compared to the corresponding KCs without affect involvement.}

\begin{table*}[h!]
    \centering
    \caption{\textbf{The impact of each component on the DASKT model.}}
    \scriptsize
    \resizebox{0.9\textwidth}{!}{%
    \begin{tabular}{lcccccccc}
        \toprule
        \multirow{2}{*}{\parbox{3.2cm}{\raggedright Model}} & \multicolumn{4}{c}{ASSIST2012} & \multicolumn{4}{c}{ASSISTchall} \\
        \cmidrule (lr){2-5} \cmidrule (lr){6-9}
        & RMSE & ACC & AUC & $r^2$ & RMSE & ACC & AUC  & $r^2$ \\
        \midrule
        \parbox[t]{3.2cm}{\raggedright DASKT}  & \textbf{0.3849} & \textbf{0.7813} & \textbf{0.7925} & \textbf{0.2248} &\textbf{0.3926} &\textbf{0.8215} &\textbf{0.8529} &\textbf{0.3246} \\
        \parbox[t]{3.2cm}{\raggedright DASKT w/o A-GAT} & 0.4043 & 0.7628 &0.7781 & 0.2109 &0.4125 &0.7946 & 0.8142 & 0.2472 \\
        \parbox[t]{3.2cm}{\raggedright DASKT w/o AT-GAT} & 0.4115 & 0.7426 &0.7518 &0.2048 &0.4189  &0.7647  &0.7798 &0.2346 \\
        \parbox[t]{3.2cm}{\raggedright DASKT w/o ICA} &0.4219 & 0.7401 & 0.7485& 0.2001&0.4215 &0.7482  &0.7682 &0.2274 \\
        \parbox[t]{3.2cm}{\raggedright DASKT w/o MAF} &0.4221 & 0.7387 &0.7404 &0.1857 &0.4249  &0.7358 &0.7671 & 0.1845
        \\ \bottomrule
    \end{tabular}
    \label{table5}}
\end{table*}

\subsection{Ablation Study (RQ4)}
In this section, we will conduct a detailed ablation study on the DASKT model to delve into the specific impact of each module on the model's performance. We will conduct experiments on four model variants, each of which removes a key component of the model. Here are detailed descriptions of the variants:
\begin{itemize}
	\item \textbf{DASKT w/o A-GAT} does not take into account the affect graph attention network. Instead, it uses the results of the affective trajectory graph attention Network as the affect for the final time step.
	\item \textbf{DASKT w/o AT-GAT} does not take into account the affective trajectory graph attention network, including not considering the A-GAT network.
	\item \textbf{DASKT w/o ICA} does not consider interval computation of affect; instead, it only uses affective factors.
	\item \textbf{DASKT w/o MAF} does not use affective factors, that is, the model does not include any affect-related data.
 \end{itemize}

Table \ref{table5} shows the experimental results of the aforementioned variant models. From the data, we can clearly see that when the A-GAT module is removed, that is, only the impact of the affect trajectory is considered without in-depth processing of the affect, the performance of the model will significantly decrease. This finding strongly supports our initial intention of modeling knowledge states under affective  states. Further observation found that the DASKT w/o AT-GAT model, which does not use dynamic affect, has an even more noticeable performance drop. This result once again validates the importance of dynamic affect in evaluating student affect coherence, emphasizing the key role of dynamic affect in student affect assessment. For the DASKT w/o ICA variant model that only considers the impact of affective factors and does not specifically process affective factors, we found that it has certain advantages compared to the DASKT w/o MAF variant model that does not consider affective factors at all. This finding indicates that refining affective factors has a positive impact on students' mastery of knowledge states during the exercise process, which also provides us with valuable guidance. Lastly, we observed that when the model does not use any affective factors, the model loses the main dependency for learning knowledge states modeling under affective  states, which has a significant negative impact on the model, preventing the model from fully predicting students' future performance. This result further emphasizes the importance of affective factors in knowledge states modeling and provides important insights for our research.

\subsection{Visualizing Future Predictions under the Influence of Affective States (RQ5)}

\begin{figure*}[!htb]
    \centering
    \includegraphics[width=1.0\textwidth]{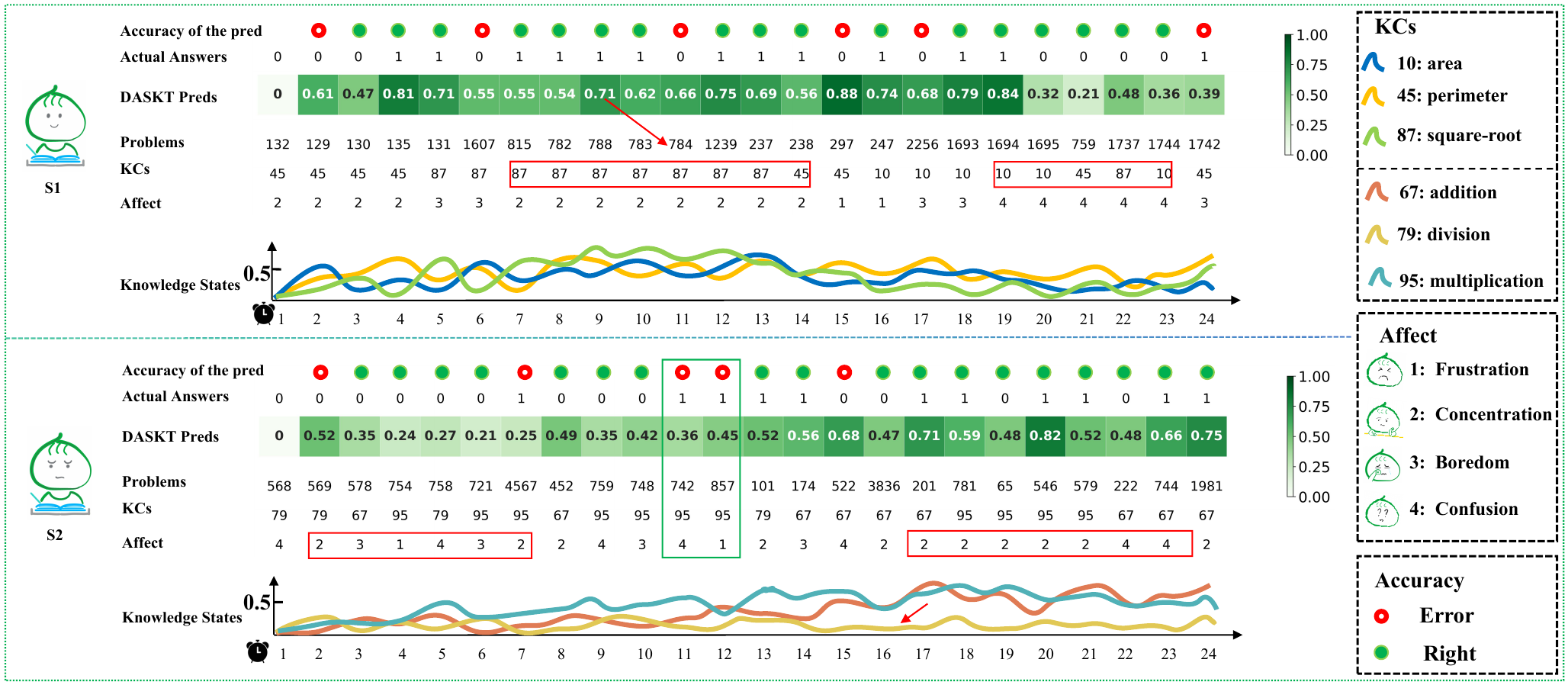}
    \caption{{\color{blue}\textbf{This is an example demonstrated on the ASSISTchall dataset, revealing the impact of dynamic affective  states on the changes in knowledge states and the performance of the DASKT model. We interpret future predictions by comparing the dynamic affective  states in the learning sequences of two students.  Accuracy of the pred describes the predictive accuracy of the DASKT model for the exercise problem at time  t. Solid circle symbols indicate that the prediction matches the student's actual answer, while hollow circle symbols show that the prediction is inconsistent with the student's actual response.}\label{figurecase}}}
\end{figure*}

In this study, we used the performances of two students from the ASSISTchall dataset to explore the accuracy of assessing student knowledge states by simulating dynamic affective  states. As shown in Figure \ref{figurecase}, the DASKT model demonstrates significant advantages in prediction performance and ability to interpret future performances in the student learning process.

Taking student $s1$ as an example, his/her affective  states throughout the learning process is relatively stable, primarily showing a state of concentration, which is indicative of a good learning state. From time step 7 to 14, the student's affect remains concentration, and the knowledge states during this period also shows a high level. For KC of 87 (square roots), the student's answers during this period are mostly correct, and the model can accurately predict the student's performance. However, at time step 11, we observed an interesting phenomenon. Despite the student correctly answering this KC recently, the student's answer on this KC turned out to be incorrect. Further analysis revealed that the interval between this question and the previous one exceeded one month. Although according to the dynamic affect design principles of our model, the student's affect is still assessed as focused, the student's affect may have changed at this moment, possibly due to poor condition or mistakes. This reminds us that the simulation calculation of affect needs more in-depth cognitive research in the future.

As the student's exercises deepen, student $s1$ begins to show confusion affect, which is a common phenomenon in the learning process. The model accurately captures this affect change and adjusts all knowledge states of the three involved KCs according to the student's performance, which aligns with our cognition. When a person is in a lower affect, their understanding of anything will decline to varying degrees.

Observing the overall affect performance of student $s2$, we found that the student struggled with their affect in the early stage and was extremely unstable. Such affect had a negative impact on the learning state. For example, during time steps 2 to 7, the student's affect changed dramatically, and the answers to the questions were mostly wrong; at this moment, the knowledge states under the affect  states also showed a very low situation. However, we also found that affect assessment cannot fully simulate human psychological changes. For example, at time steps 11 and 12, student $s2$, due to poor and unstable negative affect performance in the early stage, even if the current question corresponding to KC of 95 (multiplication) is answered correctly, the change in the knowledge states is only slightly improved, which is not enough for the DASKT model to believe that the student can correctly answer this type of question.

As the exercises deepen, during time steps 16 to 24, student $s2$'s affect performance tends to be positive, and they start to concentrate on completing the exercises. The improvement of the student's state leads to an increase in the correctness of answering questions, and the involved KCs of addition and multiplication also show significant growth. However, for the mastery of KC of 79 (division), even under positive affect performance, there is no significant improvement. The possible reason is that the student did not involve this KC during the time when they showed positive performance, so the state of this KC has been low, which also verifies that the DASKT model will not blindly modify the state of the KC.

These cases reveal the importance of knowledge tracing under affective  states. Correctly assessing affective  states has far-reaching implications for accurately assessing students' knowledge states. These findings emphasize the importance of affect assessment in the field of educational technology, especially in the context of personalized learning and prediction of student knowledge states.

\section{Conclusions and Limitations}

In this study, we explored the importance of affect assessment in students' problem-solving and introduced the DASKT model. This model enhances KT by integrating affect-enhanced student interaction features. We classified students' affective factors, such as confidence, interest, and effort, and extracted relevant data from non-affect-oriented behavioral records to compute these factors. We then used clustering and spatial sequence models to simulate dynamic changes in students' affect, ensuring continuity. By combining this affective information with the DLKT model, we enhanced its reasoning capabilities. Extensive experiments validated the model's effectiveness and interpretability, demonstrating the value of incorporating affect into knowledge states.

{\color{blue}We believe that the dynamic affect simulation method, based on educational psychology, could be extended to other areas of smart education, such as cognitive diagnosis. However, this method has not yet been evaluated outside of KT tasks. Moreover, simulating affective factors requires behavioral data, which is missing in some datasets, limiting our experiments. The lack of gold-standard affective data also hinders deeper comparisons, presenting challenges for future research.}


\bibliographystyle{IEEEtran} 
\bibliography{IEEEabrv,DASKT.bib}

\end{document}